# On the symbol error probability of regular polytopes

Erik Agrell and Magnus Karlsson

*Abstract*—An exact expression for the symbol error probability of the four-dimensional 24-cell in Gaussian noise is derived. Corresponding expressions for other regular convex polytopes are summarized. Numerically stable versions of these error probabilities are also obtained.

## I. INTRODUCTION

THE 24-CELL is a regular convex polytope (RCP) with 24 facets and 24 vertices in four dimensions. Its vertices were considered as a signal constellation for digital modulation already in 1977 [1]. The symbol error probability (SEP) of this modulation format was approximated by a union bound in [1] and the bit error probability was estimated by simulation in [2]. The constellation was considered for satellite communications in [2], [3] and has recently received renewed interest in fiber-optical communications [4], [5]. This new interest was motivated by the rapid progress in coherent detection, which makes it possible to transmit data in both polarizations simultaneously and thus to employ four-dimensional constellations [6]. In this paper, we calculate its exact SEP over the additive white Gaussian noise (AWGN) channel, which has not been done before.

We put the result into context by briefly summarizing what is known about signal constellations constructed from the vertices of other RCPs in $n$ dimensions. An RCP is a convex polytope with the maximum degree of symmetry; its facets and vertices all look the same, and the facets are in turn RCPs. Therefore, all symbols have the same energy and the same conditional error probability, which is tractable at least from a theoretical viewpoint [7]. A drawback is that the number of vertices $M$ of an RCP is usually not a power of two, but such constellations can nevertheless be used in digital communications if a block of bits is mapped to more than one symbol [8], [9], in coded modulation [10], or to reduce the peak power in orthogonal frequency-division modulation systems [11]. Furthermore, in a data communication scenario, some levels beyond a power of two are useful for framing and control purposes; indeed, 3- and 5-level modulations are already applied in standards for Fast Ethernet and Gigabit Ethernet [12, pp. 285–289].

The SEP of RCPs can be approximated by sphere packing [7], random coding [7], simplex packing [13], and union bounds [14, Sec. 3.2.7]. To obtain an exact expression for the SEP, one needs to integrate an $n$-dimensional Gaussian density, centered at an RCP vertex, over its Voronoi cell. For a given RCP, the Voronoi cells are unbounded $n$-dimensional pyramids with apex at the origin, whose $(n-1)$-dimensional base cells are the facets of the *dual* polytope. Thus, the exact SEP is in general an $n$-dimensional integral, but as we shall see, it can in most cases be simplfied into a one-dimensional integral, thanks to the inherent symmetries of RCPs.

There exist infinitely many RCPs in dimension $n = 2$, five for $n = 3$, six for $n = 4$, and three for $n \geq 5$ [15, p. 412]. For most of these constellations, the corresponding SEP is known, but at high signal-to-noise ratios, the published expressions often consist of the difference between two almost equal entities, which makes them numerically unstable. We therefore derive equivalent expressions that lend themselves better to numerical evaluation.

## II. POWER AND SPECTRAL EFFICIENCY

We define the signal-to-noise ratios $\gamma = E_s/N_0$ and $\gamma_b = E_b/N_0$, where $E_s$ is the symbol energy, $E_b = E_s/\log_2 M$ is the bit energy, $M$ is the number of levels, and $N_0/2$ is the double-sided spectral density of the AWGN. For modulation using the vertices of any RCP, the SEP $P$ can be upperbounded by the *union bound*

$$P \leq AQ\left(\sqrt{\frac{d^2}{2E_s}\gamma}\right), \qquad (1)$$

where $Q(x) \triangleq \int_x^\infty f(x)dx$, $f(x)$ is the normalized Gaussian probability density function $f(x) \triangleq (1/\sqrt{2\pi})\exp(-x^2/2)$, $d$ is the minimum Euclidean distance between pairs of vertices, and $A$ is the number of such pairs involving a given vertex.[1] The bound is an accurate approximation at medium to high $\gamma$ and it is asymptotically exact. We define the *power efficiency* as $G_b = d^2/(4E_b)$, which is the reduction of bit energy $E_b$ that a particular constellation admits, compared with binary phase-shift keying (BPSK) modulation, to attain the same, asymptotically low, SEP. Furthermore, the *spectral efficiency* $S = \log_2 M/n$ is the reduction of bandwidth that a particular constellation admits, compared with BPSK, with the same bit rate and pulse shape.

The parameters $M$, $A$, and $d$ of all RCPs are given in Tab. I, adapted from [16, pp. 292–295][2], along with $G_b$ and $S$.

## III. GENERIC RCPs

Three types of generic RCPs exist for all $n \geq 2$. First, the $n$-cube can be defined as the vertex set $(\pm 1, \ldots, \pm 1)$, with all $2^n$ sign combinations. When used for modulation, this constellation results in independent BPSK modulation in all $n$ dimensions. It specializes into a square and cube for

---

E. Agrell is with the Dept. of Signals and Systems, Chalmers Univ. of Technology, SE-41296 Göteborg, Sweden, email agrell@chalmers.se. M. Karlsson is with the Dept. of Microtechnology and Nanoscience, Chalmers Univ. of Technology, SE-41296 Göteborg, Sweden.

[1]The inequality does however not hold for irregular polytopes, for which in general not only the closest neighbors contribute to the Voronoi cell.

[2]The parameters were calculated from $N_0$, $N_1$, and $\phi$ in [16] as $M = N_0$, $A = 2N_1/N_0$, and $d/\sqrt{E_s} = 2\sin\phi$.

$n = 2$ and $n = 3$, resp. Second, the $n$-crosspolytope (or orthoplex), whose $2n$ vertices can be taken as all permutations of $(\pm 1, 0, \ldots, 0)$, is the geometric dual of the $n$-cube and corresponds to biorthogonal modulation. It is also a square for $n = 2$ but an octahedron for $n = 3$. The third type is the self-dual $n$-simplex, which is an equilateral triangle for $n = 2$ and a regular tetrahedron for $n = 3$. Its $n+1$ vertices can be taken as $(1-\alpha, \ldots, 1-\alpha)$ and all permutations of $(\alpha - n, 1, \ldots, 1)$, where $\alpha = \sqrt{n+1}$ and all vectors have length $n$.[3]

The SEP of these three RCPs are given by the following theorem, where two expressions are given for each SEP; one that is suitable for analysis and another for numerical evaluation. Expressions related to our (5) and (7), but less reliable at very high $\gamma$, were given in [17].

*Theorem 1:* The SEPs of the $n$-cube $P_\text{c}$, $n$-crosspolytope $P_\text{cp}$, and $n$-simplex $P_\text{s}$ are

$$P_\text{c} = 1 - \left[1 - Q\left(\sqrt{\frac{2\gamma}{n}}\right)\right]^n \quad (2)$$

$$= Q\left(\sqrt{\frac{2\gamma}{n}}\right) \sum_{i=0}^{n-1} \binom{n}{i+1} \left[-Q\left(\sqrt{\frac{2\gamma}{n}}\right)\right]^i, \quad (3)$$

$$P_\text{cp} = 1 - \frac{1}{\sqrt{2\pi}} \int_0^\infty (1 - 2Q(x))^{n-1} \exp\left(-\frac{(x - \sqrt{2\gamma})^2}{2}\right) dx \quad (4)$$

$$= Q(\sqrt{2\gamma}) + \frac{2}{\sqrt{\pi}} \int_0^\infty Q(x) \exp\left(-\frac{(x - \sqrt{2\gamma})^2}{2}\right)$$
$$\cdot \sum_{i=0}^{n-2} \binom{n-1}{i+1} (-2Q(x))^i dx, \quad (5)$$

$$P_\text{s} = 1 - \frac{1}{\sqrt{2\pi}} \int_{-\infty}^\infty (1 - Q(x))^n$$
$$\cdot \exp\left[-\frac{1}{2}\left(x - \sqrt{\frac{2\gamma(n+1)}{n}}\right)^2\right] dx \quad (6)$$

$$= \frac{1}{\sqrt{2\pi}} \int_{-\infty}^\infty Q(x) \exp\left[-\frac{1}{2}\left(x - \sqrt{\frac{2\gamma(n+1)}{n}}\right)^2\right]$$
$$\cdot \sum_{i=0}^{n-1} \binom{n}{i+1} (-Q(x))^i dx. \quad (7)$$

*Proof:* Transmission over the $n$-cube is equivalent to transmitting a block of $n$ bits over independent BPSK channels, which yields (2). Expressions (4) and (6) were derived from [14, eqs. (4.102) and (4.116)] via the substitutions $q = x/\sqrt{2} - \sqrt{\gamma}$ and $q = x/\sqrt{2} - \sqrt{\gamma(1+1/n)}$, resp. Expressions (3), (5), and (7) follow after expanding $(1-Q(\cdot))^n$ by the binomial theorem and integrating out the constant term. □

Other geometric properties of the three generic RCPs were calculated in [18].

---
[3]To our best knowledge, this is the first time that an explicit $n$-dimensional coordinate representation is presented for the $n$-simplex. ($\alpha = -\sqrt{n+1}$ gives another, equally valid, representation.) The standard representation, which is in $n+1$ dimensions, is less useful for transmission.

TABLE I
PARAMETERS OF ALL RCPS.

| RCP | $n$ | $M$ | $A$ | $d/\sqrt{E_\text{s}}$ | $G_\text{b}$ (dB) | $S$ |
|---|---|---|---|---|---|---|
| $M$-polygon | 2 | $M$ | 2 | $2\sin\frac{\pi}{M}$ | | |
| dodecahedron | 3 | 20 | 3 | $\frac{\sqrt{5}-1}{\sqrt{3}}$ | –2.59 | 1.44 |
| icosahedron | 3 | 12 | 5 | $\sqrt{2 - \frac{2}{\sqrt{5}}}$ | –0.04 | 1.19 |
| 24-cell | 4 | 24 | 8 | 1 | 0.59 | 1.15 |
| 120-cell | 4 | 600 | 4 | $\frac{3-\sqrt{5}}{2\sqrt{2}}$ | –7.73 | 2.31 |
| 600-cell | 4 | 120 | 12 | $\frac{\sqrt{5}-1}{2}$ | –1.81 | 1.73 |
| $n$-cube | $n$ | $2^n$ | $n$ | $\frac{2}{\sqrt{n}}$ | 0 | 1 |
| $n$-crosspolytope | $n$ | $2n$ | $2(n-1)$ | $\sqrt{2}$ | | |
| $n$-simplex | $n$ | $n+1$ | $n$ | $\sqrt{2 + \frac{2}{n}}$ | | |

## IV. TWO-DIMENSIONAL RCPS

In two dimensions, RCPs are polygons, which exist with an arbitrary number of vertices $M \geq 3$. Using such a constellation for transmission corresponds to $M$-ary phase shift keying ($M$-PSK). The following expression, which was given in [19], is both simple and stable.

*Theorem 2:* The SEP of the $M$-polygon $P_\text{PSK}$ is

$$P_\text{PSK} = \frac{1}{\pi} \int_0^{\pi-\pi/M} \exp\left[-\gamma\left(\frac{\sin(\pi/M)}{\sin z}\right)^2\right] dz.$$

The SEP for selected values of $M$ is plotted in Fig. 1 (a). As expected, $M = 4$ has the same asymptotic performance as BPSK. The only $M$-PSK format with a power efficiency higher than 0 dB is 3-PSK, as shown already in [20].

## V. THREE-DIMENSIONAL RCPS

In three dimensions, there are two RCPs in addition to the above-mentioned cube, octahedron, and tetrahedron. These are the icosahedron and the dodecahedron, with 12 and 20 vertices, respectively, which are duals of each other. Their exact SEPs are not known, but their union bounds follow from (1) and Table I. The results are shown in Fig. 1 (b). The tetrahedron (3-simplex) has the highest power efficiency of the three-dimensional RCPs [5].

## VI. FOUR-DIMENSIONAL RCPS

Coxeter remarked that a "peculiarity of four-dimensional space is the occurrence of the 24-cell..., having no analogue above or below" [16, p. 289]. The 24-cell is a four-dimensional self-dual RCP with 24 vertices. It is the only RCP, in any dimension, that is both more power efficient and more spectrally efficient than the $n$-cube (BPSK modulation). Curiously enough, the power and spectral efficiencies of the 24-cell are equal, $d^2/(4E_\text{b}) = \log_2 M/n = (3 + \log_2 3)/4 = 1.15$.

The 24-cell can be described in several ways depending on its orientation. The vertices can be taken as the union of $(\pm 1, \pm 1, \pm 1, \pm 1)$, with all possible combinations of signs, and $\Pi(2, 0, 0, 0)$, where $\Pi(\cdot)$ means the set of all permutations of the given elements. In other words, the 24-cell is the convex hull of the union of the 4-cube and the 4-crosspolytope, scaled to the same circumscribed radius and properly aligned. This description, although geometrically appealing, does not seem to easily admit an exact computation of the SEP.

Rotating the constellation, the vertices of the 24-cell can be described as the set

$$\mathcal{X} \triangleq \Pi(\pm 1, \pm 1, 0, 0),$$

with all possible permutations and signs. With this coordinate representation, we are able to calculate the SEP of the 24-cell. The result is again presented in the form of two expressions, where the first is analytically simpler and the second numerically useful even at high $\gamma$ values.

*Theorem 3:* The SEP of the 24-cell is

$$P_{24} = 1 - \sqrt{\frac{2}{\pi}} \int_0^\infty \exp\left(-\frac{(x-\sqrt{\gamma})^2}{2}\right)$$
$$\cdot (1 - 2Q(x))^2 Q(x - \sqrt{\gamma}) dx \quad (8)$$
$$= Q(\sqrt{\gamma})(2 - Q(\sqrt{\gamma})) + \sqrt{\frac{32}{\pi}} \int_0^\infty \exp\left(-\frac{(x-\sqrt{\gamma})^2}{2}\right)$$
$$\cdot Q(x)(1 - Q(x))Q(x - \sqrt{\gamma}) dx. \quad (9)$$

*Proof:* A signal vector $\boldsymbol{X}\sqrt{E_s/2}$, where $\boldsymbol{X}$ is randomly taken from $\mathcal{X}$ with equal probabilities, is transmitted over a discrete-time AWGN channel with variance $N_0/2$. The received vector is $\boldsymbol{X}\sqrt{E_s/2} + \boldsymbol{Z}\sqrt{N_0/2}$, where $\boldsymbol{Z} = (Z_1, Z_2, Z_3, Z_4)$ is a vector of independent, zero-mean, unit-variance, Gaussian random variables. To simplify the analysis, the received vector is rescaled by a factor $\sqrt{E_s/2}$. Thus, the maximum likelihood detector shall find the vector in $\mathcal{X}$ being closest to $\boldsymbol{X} + \boldsymbol{Z}/\sqrt{\gamma}$ in the Euclidean sense.

Because the signal vectors are equally probable and symmetrically located, we assume, without loss of generality, that the transmitted vector is $\boldsymbol{x}_1 \triangleq (1, 1, 0, 0)$. The SEP can now be expressed as

$$P_{24} = \Pr\{\boldsymbol{x}_1 + \frac{1}{\sqrt{\gamma}}\boldsymbol{Z} \notin \Omega_{\mathcal{X}}(\boldsymbol{x}_1)\}, \quad (10)$$

where $\Omega_{\mathcal{X}}(\boldsymbol{u})$ for any $\boldsymbol{u} \in \mathbb{R}^4$ is the Voronoi cell[4]

$$\Omega_{\mathcal{X}}(\boldsymbol{u}) \triangleq \{\boldsymbol{y} \in \mathbb{R}^4 \mid \|\boldsymbol{y} - \boldsymbol{u}\| \leq \|\boldsymbol{y} - \boldsymbol{x}\|, \boldsymbol{x} \in \mathcal{X}\}. \quad (11)$$

In order to integrate the four-dimensional noise over the Voronoi cell, we need to find a compact coordinate representation for this cell. To this end, we first partition $\mathcal{X}$ into the subsets

$\mathcal{X}_1 \triangleq \{(1, 1, 0, 0)\},$
$\mathcal{X}_2 \triangleq \{(0, 1, 0, \pm 1), (0, 1, \pm 1, 0), (1, 0, 0, \pm 1), (1, 0, \pm 1, 0)\},$
$\mathcal{X}_3 \triangleq \{(-1, 1, 0, 0), (1, -1, 0, 0)\},$
$\mathcal{X}_4 \triangleq \{(0, 0, \pm 1, \pm 1)\},$
$\mathcal{X}_5 \triangleq \{(0, -1, 0, \pm 1), (0, -1, \pm 1, 0),$
$\qquad (-1, 0, 0, \pm 1), (-1, 0, \pm 1, 0)\},$
$\mathcal{X}_6 \triangleq \{(-1, -1, 0, 0)\}.$

[4]In our notation, the definition is valid also when $\boldsymbol{u} \notin \mathcal{X}$, which will be utilized in (12)–(17).

With these definitions,

$$\mathcal{X} = \bigcup_{i=1}^6 \mathcal{X}_i,$$
$$\Omega_{\mathcal{X}}(\boldsymbol{u}) = \bigcap_{i=1}^6 \Omega_{\mathcal{X}_i}(\boldsymbol{u}), \text{ for any } \boldsymbol{u} \in \mathbb{R}^4. \quad (12)$$

Applying (11) to the subset $\mathcal{X}_2$ yields

$\Omega_{\mathcal{X}_2}(\boldsymbol{x}_1)$
$= \{\boldsymbol{y} \in \mathbb{R}^4 \mid \|\boldsymbol{y} - \boldsymbol{x}_1\|^2 \leq \|\boldsymbol{y} - \boldsymbol{x}\|^2, \boldsymbol{x} \in \mathcal{X}_2\}$
$= \{\boldsymbol{y} \in \mathbb{R}^4 \mid \langle \boldsymbol{y}, \boldsymbol{x}_1 - \boldsymbol{x}\rangle \geq 0, \boldsymbol{x} \in \mathcal{X}_2\}$
$= \{\boldsymbol{y} \in \mathbb{R}^4 \mid y_1 \geq |y_4|, y_1 \geq |y_3|, y_2 \geq |y_4|, y_2 \geq |y_3|\}$
$= \{\boldsymbol{y} \in \mathbb{R}^4 \mid \min\{y_1, y_2\} \geq \max\{|y_3|, |y_4|\}\}, \quad (13)$

where $\boldsymbol{y} = (y_1, y_2, y_3, y_4)$ and $\langle \cdot, \cdot \rangle$ denotes the inner product. The Voronoi cells of $\boldsymbol{x}_1$ with respect to the subsets $\mathcal{X}_3, \ldots, \mathcal{X}_6$ can be calculated by the same method. Omitting the details, the results are

$$\Omega_{\mathcal{X}_3}(\boldsymbol{x}_1) = \{\boldsymbol{y} \in \mathbb{R}^4 \mid \min\{y_1, y_2\} \geq 0\} \quad (14)$$
$$\Omega_{\mathcal{X}_4}(\boldsymbol{x}_1) = \{\boldsymbol{y} \in \mathbb{R}^4 \mid y_1 + y_2 \geq |y_3| + |y_4|\} \quad (15)$$
$$\Omega_{\mathcal{X}_5}(\boldsymbol{x}_1) = \{\boldsymbol{y} \in \mathbb{R}^4 \mid y_1 + y_2 + \min\{y_1, y_2\}$$
$$\geq \max\{|y_3|, |y_4|\}\} \quad (16)$$
$$\Omega_{\mathcal{X}_6}(\boldsymbol{x}_1) = \{\boldsymbol{y} \in \mathbb{R}^4 \mid y_1 + y_2 \geq 0\}. \quad (17)$$

Comparing (13) with (14) yields $\Omega_{\mathcal{X}_2}(\boldsymbol{x}_1) \subseteq \Omega_{\mathcal{X}_3}(\boldsymbol{x}_1)$, because $\min\{y_1, y_2\} \geq \max\{|y_3|, |y_4|\} \Rightarrow \min\{y_1, y_2\} \geq 0$. Similarly, comparing (13) with each of (15)–(17) shows that $\Omega_{\mathcal{X}_2}(\boldsymbol{x}_1) \subseteq \Omega_{\mathcal{X}_i}(\boldsymbol{x}_1)$ for $i = 4, 5, 6$. Now using these results in (12) yields

$$\Omega_{\mathcal{X}}(\boldsymbol{x}_1) = \Omega_{\mathcal{X}_2}(\boldsymbol{x}_1). \quad (18)$$

We can now finish the SEP calculation. From (10), (13), and (18),

$$P_{24} = 1 - \Pr\left\{\boldsymbol{x}_1 + \frac{1}{\sqrt{\gamma}}\boldsymbol{Z} \in \Omega_{\mathcal{X}_2}(\boldsymbol{x}_1)\right\}$$
$$= 1 - \Pr\left\{\min\left\{1 + \frac{Z_1}{\sqrt{\gamma}}, 1 + \frac{Z_2}{\sqrt{\gamma}}\right\} \geq \max\left\{\frac{|Z_3|}{\sqrt{\gamma}}, \frac{|Z_4|}{\sqrt{\gamma}}\right\}\right\}$$
$$= 1 - \Pr\{\sqrt{\gamma} + \min\{Z_1, Z_2\} \geq \max\{|Z_3|, |Z_4|\}\}$$
$$= 1 - \Pr\{Z_1 \leq Z_2, \sqrt{\gamma} + \min\{Z_1, Z_2\} \geq \max\{|Z_3|, |Z_4|\}\}$$
$$\quad - \Pr\{Z_1 > Z_2, \sqrt{\gamma} + \min\{Z_1, Z_2\} \geq \max\{|Z_3|, |Z_4|\}\}.$$

By symmetry, the last two probabilities are equal; thus,

$$P_{24} = 1 - 2\Pr\{Z_1 \leq Z_2, \sqrt{\gamma} + \min\{Z_1, Z_2\} \geq \max\{|Z_3|, |Z_4|\}\}$$
$$= 1 - 2\Pr\{Z_1 \leq Z_2, \sqrt{\gamma} + Z_1 \geq |Z_3|, \sqrt{\gamma} + Z_1 \geq |Z_4|\}.$$

Finally, we marginalize the probability over $Z_1 = z_1$ to obtain

$$P_{24} = 1 - 2 \int_{-\sqrt{\gamma}}^\infty f(z_1)$$
$$\cdot \Pr\{Z_2 \geq z_1, |Z_3| \leq \sqrt{\gamma} + z_1, |Z_4| \leq \sqrt{\gamma} + z_1\} dz_1$$
$$= 1 - 2 \int_{-\sqrt{\gamma}}^\infty f(z_1) Q(z_1)(1 - 2Q(\sqrt{\gamma} + z_1))^2 dz_1 \quad (19)$$
$$= 1 - 2 \int_0^\infty f(z - \sqrt{\gamma}) Q(z - \sqrt{\gamma})(1 - 2Q(z))^2 dz,$$

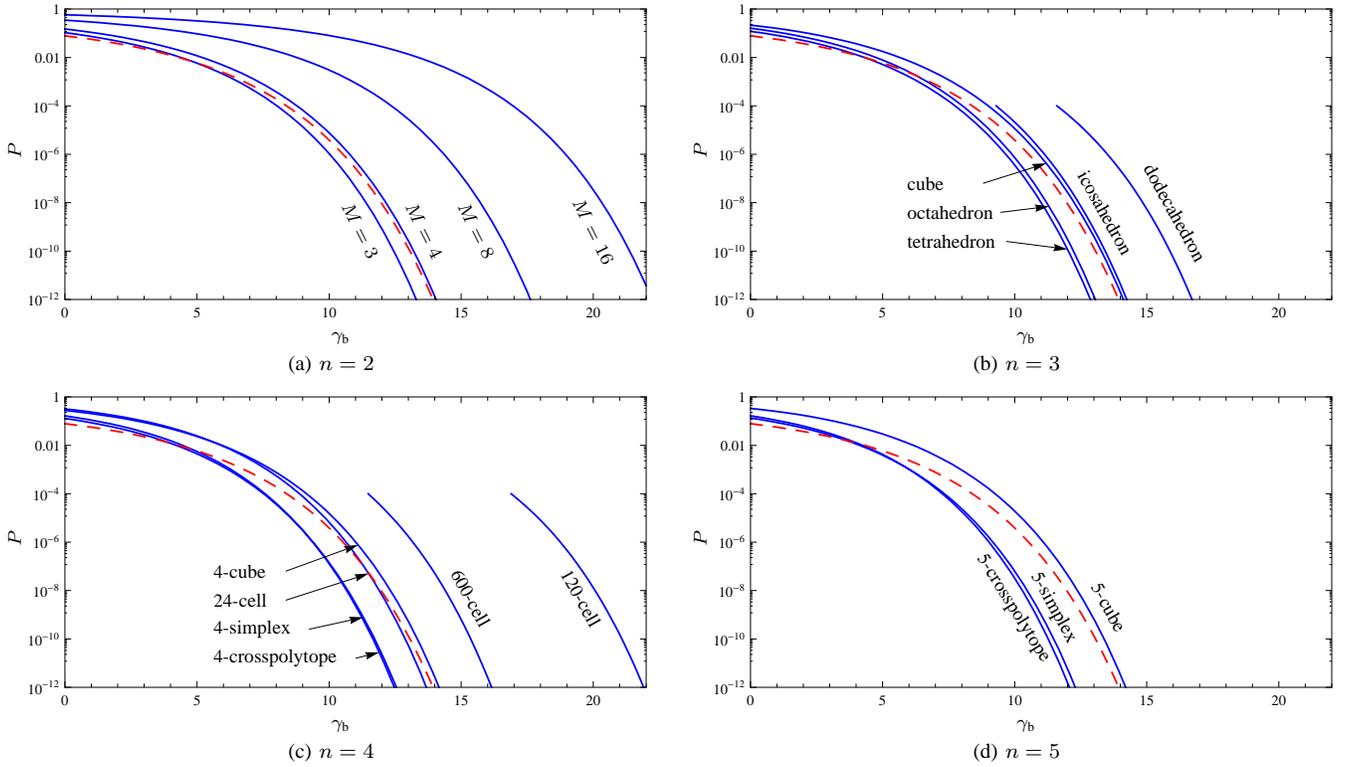

Fig. 1. The SEP of RCPs in various dimensions $n$. Except for the icosahedron, dodecahedron, 600-cell, and 120-cell, whose SEPs are approximated by the union bound, all SEPs are exact. The SEP of BPSK is shown as a reference in all diagrams (dashed line).

which completes the proof of (8).

To prove (9), we first note that for any $x \in \mathbb{R}$,

$$\begin{aligned} Q^2(x) &= \Pr\{Z_1 \geq x, Z_2 \geq x\} \\ &= 2\Pr\{Z_2 \geq Z_1 \geq x\} \\ &= 2\int_x^\infty f(z_1)Q(z_1)dz_1. \end{aligned} \quad (20)$$

Expanding (19) with the help of (20) yields

$$\begin{aligned} P_{24} &= 1 - 2\int_{-\sqrt{\gamma}}^\infty f(z_1)Q(z_1) \\ &\quad \cdot (1 - 4Q(\sqrt{\gamma}+z_1) + 4Q^2(\sqrt{\gamma}+z_1))dz_1 \\ &= 1 - Q^2(-\sqrt{\gamma}) + 2\int_{-\sqrt{\gamma}}^\infty f(z_1)Q(z_1) \\ &\quad \cdot (4Q(\sqrt{\gamma}+z_1) - 4Q^2(\sqrt{\gamma}+z_1))dz_1 \\ &= 1 - (1-Q(\sqrt{\gamma}))^2 + 8\int_{-\sqrt{\gamma}}^\infty f(z_1)Q(z_1) \\ &\quad \cdot Q(\sqrt{\gamma}+z_1)(1-Q(\sqrt{\gamma}+z_1))dz_1 \\ &= Q(\sqrt{\gamma})(2-Q(\sqrt{\gamma})) \\ &\quad + 8\int_0^\infty f(z-\sqrt{\gamma})Q(z-\sqrt{\gamma})Q(z)(1-Q(z))dz, \end{aligned}$$

which completes the proof of (9). □

The SEP of all four-dimensional RCPs is shown in Fig. 1 (c). It can be observed that $n = 4$ is the lowest dimension for which the crosspolytope has the highest power efficiency of all RCPs. In addition to the 24-cell and the three generic polytopes, there are RCPs with 120 and 600 vertices in four dimensions. Miyazaki made an artistic effort to visualize these polytopes [21, Ch. 14], which are each other's duals. Their power efficiencies are low, similarly to the two largest RCPs in three dimensions.

## VII. HIGHER-DIMENSIONAL RCPs

In dimensions $n \geq 5$, there are no RCPs apart from the three generic types. For medium and high $\gamma_b$, the crosspolytope is always better than the simplex, and the simplex is better than the cube. Their SEP performances are shown in Fig. 1 (d) for the five-dimensional case.

As the dimension increases, the power efficiency gap between the simplex and the cube increases monotonically. However, the gap between the crosspolytope and the simplex, which based on Tab. I and the definition of $G_b$ can be expressed as

$$\frac{G_{\mathrm{b,cp}}}{G_{\mathrm{b,s}}} = \frac{n\log_2(2n)}{(n+1)\log_2(n+1)}, \quad (21)$$

increases until it reaches a maximum for $n = 24$, where the crosspolytope is 0.62 dB better. Thereafter the gap decreases again and approaches 0 dB in very high dimensions. Extending $n$ to real numbers, the maximum of (21) occurs for $n \approx 24.066$, which is quite close to 24. Whether this is a pure coincidence or more deeply related to other unique properties of 24-dimensional geometry, such as the existence of the Leech lattice, is not known.